\documentclass[12pt]{article}
\usepackage{amsmath,amssymb,mathrsfs}
\usepackage{authblk}

\newtheorem{theorem}{Theorem}

\newtheorem{definition}[theorem]{Definition}

\begin{document}

\title{Quantum Stochastic Processes and the Modelling of Quantum Noise\thanks{Contributed article to the Second Edition of Springer's Encyclopedia of Systems and Control (to appear)}}

\author{Hendra I. Nurdin}      

\affil{School of Electrical Engineering and Telecommunications, University of New South Wales (UNSW), UNSW Sydney 2052, Australia. Email: h.nurdin@unsw.edu.au}

\date{}

\maketitle

\begin{abstract}
This brief article gives an overview of quantum mechanics as a {\em quantum probability theory}. It begins with a review of the basic operator-algebraic elements that connect probability theory with quantum probability theory. Then quantum stochastic processes is formulated as a generalization of stochastic processes within the framework of quantum probability theory. Quantum Markov models from quantum optics are used to explicitly illustrate the underlying abstract concepts and their connections to the quantum regression theorem from quantum optics.
\end{abstract}

\section{Introduction}

Many phenomena display dynamics that appear to be random and can often be accurately modeled as stochastic processes. 
The modern theory of  stochastic processes is built upon the measure-theoretic axiomatization of probability theory as developed by Kolmogorov in the 1930's; for a historical overview, see, e.g., \cite{Bing00,SV06}. These theories underpin the  stochastic systems theory and stochastic control theory that find wide applications in disciplines such as engineering (in particular control engineering), economics, and mathematical finance.  At the atomic  size and energy scales, where classical physics is superseded by quantum mechanics, randomness is inherent. Indeed, measurement of a quantum mechanical system yields a random outcome, as dictated by the measurement postulate of quantum mechanics. Research into the axiomatic foundation of quantum mechanics since the seminal work of von Neumann have lead to the recognition that quantum mechanics is essentially a non-commutative generalization of probability theory. That is, {\em quantum mechanics is a quantum probability theory}. 

Quantum probability theory provides a rigorous basis for studying quantum stochastic processes and developing a modern formulation of quantum stochastic systems and control theory. This facilitates the systematic formulation and solution of estimation and feedback control problems for quantum systems \cite{BvHJ07,BvH08,GJ09,NY17}. This article provides an overview of  quantum probability theory and the formulation of quantum stochastic processes and quantum Markov processes based on this theory.  The abstract mathematical concepts introduced will then be explicitly illustrated in the context of quantum Markov models for a large class of quantum optical devices.

We will use fairly standard notations, with additional  notations introduced later in the text. $\mathbb{R}$, $\mathbb{C}$ and $\mathbb{R}_{0+}$ denote the real and complex numbers, and $\mathbb{R}_{0+}=[0,\infty)$, respectively. For $c \in \mathbb{C}$, $\overline{c}$ is its complex conjugate. For a complex-valued function $X$, $\overline{X}(\cdot) = \overline{X(\cdot)}$. For a set $S$, $S^n$ denotes the $n$-fold direct product $S^n =\underbrace{S \times S \times \cdots S}_{\hbox{$n$ times}}$.  The indicator  function for a set $A$ is denoted by $\mathbf{1}_A$. A Dirac ket  $|x\rangle$ denotes a complex  vector in a Hilbert space  while a bra $\langle x |$  denotes the conjugate transpose (or dual functional) of the  vector. Thus $\langle x|y \rangle$ is the inner product of $|x\rangle$ and $|y \rangle$.  For any operator $X$ mapping a Hilbert space to another, $X^{\dag}$ denotes the adjoint of $X$ (if $X$ maps a Hilbert space to itself then $X^{\dag}$ is also referred to as the Hermitian conjugate). ${\rm Tr}(X)$ denotes the trace of a trace-class operator.  For  a self-adjoint (Hermitian) operator $X$, $X \geq 0$ denotes a positive operator, $\langle x| X | x\rangle \geq 0$ for all elements $x$ of the Hilbert space.   

\section*{Keywords}
Quantum feedback control, quantum stochastic systems, quantum stochastic control

\section{Quantum probability theory and quantum stochastic processes}


A (classical)\footnote{In this article we will use the qualifier ``classical" in brackets to emphasize that a theory is not quantum mechanical} probability space is a tuple $(\Omega,\mathcal{F},\mathbb{P})$, where $\Omega$ is the sample space, $\mathcal{F}$ the event set as a $\sigma$-algebra of subsets of $\Omega$, and $\mathbb{P}$ is a probability measure on the measure space $(\Omega,\mathcal{F})$. 
We endow $\mathbb{C}$ with its Borel $\sigma$-algebra $\mathcal{B}(\mathbb{C})$, making $(\mathbb{C},\mathcal{B}(\mathbb{C}))$  a measurable space.
A random variable $X: (\Omega, \mathcal{F}) \rightarrow (\mathbb{C},\mathcal{B}(\mathbb{C}))$ is a measurable complex-valued map from $\Omega$ to $\mathbb{C}$ (for the sake of simplicity,  we intentionally restrict ourselves only to random variables taking values in $\mathbb{C}$).  The expectation value $\mathbb{E}[X]$ of $X$ is given by $\mathbb{E}[X]=\int_{\Omega} X(\omega) P(d\omega)$. Given a sub-$\sigma$-algebra $\mathcal{G}$ of $\mathcal{F}$, the conditional expectation of a random variable $X$ on $\mathcal{G}$, denoted by $\mathbb{E}[\left. X \right| {\mathcal{G}}]$, is a random variable with the properties: (i)
$\mathbb{E}[\left. X \right| {\mathcal{G}}]$ is $\mathcal{G}$-measurable, and (ii) $\mathbb{E}[\mathbb{E}[\left. X \right| {\mathcal{G}}]K]=\mathbb{E}[XK]$ for any $\mathcal{G}$-measurable random variable $K$. It follows from this definition that if $\mathcal{G}$ and $\mathcal{H}$ are sub-$\sigma$-algebras of $\mathcal{F}$ with $\mathcal{H} \subset \mathcal{G}$ then $\mathbb{E}[\left. \mathbb{E}[\left. X \right| {\mathcal{G}}]\right| \mathcal{H}] =\mathbb{E}[\left. X \right| \mathcal{H}]$ (the tower property). 

For the rest of the article, let $T \subseteq \mathbb{R}$ denote a time index. For example, $T$ can be discrete as in $T=\{0,1,2,\ldots \}$ or it can be continuous as in $T=[0,\infty)$. Given a probability space $(\Omega,\mathcal{F},\mathbb{P})$, a classical stochastic process over  $T$ is a collection of random variables $\{X_t\}_{t \in T}$ on $(\Omega,\mathcal{F},\mathbb{P})$  indexed by elements of $T$. From a practical perspective,  one would not start with a given $(\Omega,\mathcal{F},\mathbb{P})$ but with the specification of a countable collection of finite-dimensional probability distributions $F_{t_1,t_2,\ldots,t_n}(C_1,C_2,\ldots,C_n)=\mathbb{P}(X_{t_1} \in C_1,X_{t_2} \in C_2, \ldots,X_{t_n} \in C_n)$ for any integer $n \geq 1$, any collection of distinct points $t_1,t_2,\ldots,t_n \in T$ and any $C_j \in \mathcal{B}(\mathbb{C})$, $j=1,2,\ldots,n$, which describe the stochastic process. As is well-known, under the Kolmogorov consistency conditions on the finite-dimensional distributions, a probability space $(\Omega,\mathcal{F},\mathbb{P})$ can be reconstructed on which the stochastic process $\{X_t\}_{t \in T}$ satisfying all the specifications of the finite-dimensional distributions can be realized; see, e.g., \cite{Bil86}.

An equivalent formulation of probability theory in terms of algebras of operators over a Hilbert space, attributed to Gelfand \cite[Theorem 3.2]{Streat00}, is the key to connecting probability theory to quantum mechanics. Consider an essentially bounded random variable $X$ on $(\Omega,\mathcal{F},\mathbb{P})$, $\mathop{{\rm ess} \sup}_{\omega \in \Omega} |X(\omega)| <\infty$. Denote the class of all such random variables by $L^{\infty}(\Omega, \mathcal{F},\mathbb{P})$. Each element of $L^{\infty}(\Omega, \mathcal{F},\mathbb{P})$  may be viewed as a multiplication operator  on elements of the Hilbert space  $L^2(\Omega, \mathcal{F},\mathbb{P})$ of square-integrable complex-valued functions on $(\Omega,\mathcal{F},\mathbb{P})$, in the sense that $X: Y(\cdot)  \in L^2(\Omega,\mathcal{F},\mathbb{P}) \mapsto X(\cdot)Y(\cdot) \in  L^2(\Omega,\mathcal{F},\mathbb{P})$. The class $L^{\infty}(\Omega, \mathcal{F},\mathbb{P})$ is closed under the `+' operation ($X_1,X_2 \in L^{\infty}(\Omega, \mathcal{F},\mathbb{P}) \rightarrow X_1+X_2 \in  L^{\infty}(\Omega, \mathcal{F},\mathbb{P})$) and under the commutative operator composition operation ($X_1,X_2 \in L^{\infty}(\Omega, \mathcal{F},\mathbb{P}) \rightarrow X_1 X_2 =X_2 X_1 \in  L^{\infty}(\Omega, \mathcal{F},\mathbb{P})$). Moreover, an antilinear  unary involution operator can be defined on  $L^{\infty}(\Omega, \mathcal{F},\mathbb{P})$ as  $(^*):  X \mapsto \overline{X}$. Thus, $L^{\infty}(\Omega, \mathcal{F},\mathbb{P})$ forms a commutative *-algebra of operators. 
 
A state $\mu$ on a *-operator algebra $\mathscr{A}$ with identity operator  $I_{\mathscr{A}}$ (with respect to the composition operation) is a linear complex functional on $\mathscr{A}$ which is positive (i.e., $\mu(A) \geq 0$ for all $0 \leq A \in \mathscr{A}$) and unital, $\mu(I_{\mathscr{A}})=1$.  A  map $M$ acting on $\mathscr{A}$ (such as the state $\mu$) is said to be normal if  $M(\sup_{\alpha} A_{\alpha})=\sup_{\alpha}M(A_{\alpha})$ for any bounded increasing net $\{A_{\alpha}\}$ of positive elements in $\mathscr{A}$. The algebra $L^{\infty}(\Omega, \mathcal{F},\mathbb{P})$ is a commutative   *-operator algebra on $L^2(\Omega, \mathcal{F},\mathbb{P})$ (since the composition operation is commutative) and $\mathbb{E}[\cdot]$  is a normal state on this algebra; see, e.g., \cite{BvHJ07} and the references therein. More precisely,  the *-algebra is a commutative von Neumann algebra \cite{BR79,BvHJ07}.
 
Conversely, it can be shown that any commutative  von Neumann algebra $\mathscr{A}$  and a unital normal state $\mu$ on $\mathscr{A}$ is *-isomorphic (a bijective correspondence that preserves the involution operation $^*$) to $(L^{\infty}(\Omega,\mathcal{F},\nu), \mathbb{E})$, where  $\mathbb{E}[\cdot]=\int_{\Omega} (\cdot)(\omega)\mathbb{P}(d\omega)$ for some probability measure $\mathbb{P}$ on $(\Omega,\mathcal{F})$ which is absolutely continuous with respect to $\nu$. This suggests that a natural generalization of probability theory would be a non-commutative von Neumann algebra $\mathscr{A}$ with identity operator $I_{\mathscr{A}}$ that is endowed with a unital normal state $\mu$; see \cite{Streat00} for a  historical  overview of quantum probability theory.   The pair $(\mathscr{A},\mu)$  is referred to as a {\em quantum probability space}. If $\mathscr{X} \subset \mathscr{A}$ is a Neumann algebra containing $I_{\mathscr{A}}$ then a positive linear map $E^{\mathscr{X}}: \mathscr{A} \rightarrow \mathscr{X}$ is a conditional expectation map if $\mu(x_1Ax_2) =\mu(x_1E^{\mathscr{X}}[A]x_2)$ for any $A \in \mathscr{A}$ and $x_1,x_2 \in \mathscr{X}$.

The physical interpretation associated with a quantum probability space is as follows.  The underlying Hilbert space of the von Neumann algebra corresponds  to the Hilbert space of a quantum mechanical system. The  Hermitian elements of the algebra represent physical observables. Orthogonal projection operators in the algebra (operators $P$ satisfying $P^2=P=P^{\dag}$) represent events that can take place, such as the event that ``Observable $X$ takes on the value $c$ on measurement". Two events $P_1$ and $P_2$ are compatible if they commute; these are events that can be assigned a joint probability distribution in the classical sense. 
If the initial state is the density operator $\rho$, then $\mu(X)={\rm Tr}(\rho X)$ for any observable $X$. As will be seen next, in a quantum probability space it is more natural to consider quantum dynamics in the Heisenberg picture, where operators evolve in time and the state is fixed to the initial one. 


Let $\mathscr{B}$ and $\mathscr{A}$ be von Neumann algebras with identity operators $I_{\mathscr{A}}$ and $I_{\mathscr{B}}$, respectively, and $(\mathscr{A},\mu)$ a quantum probability space. A {\em quantum stochastic process} over $\mathscr{B}$ indexed by $T \subseteq \mathbb{R}$ is a tuple $(\mathscr{A},\{j_t\}_{t\in T},\mu)$ where $j_t$ is a *-homomorphism of $\mathscr{B}$ into $\mathscr{A}$ for each $t \in T$ (i.e., $j_t(A^{\dag})=j_t(A)^{\dag}$ and $j_t(AB)=j_t(A)j_t(B)$ for any $A,B \in \mathscr{A}$) such that $j_t(I_{\mathscr{B}})=I_{\mathscr{A}}$,  and $\mathscr{A}={\rm vN}(\{j_t(\mathscr{B}),\,  t \in T\})$. Here $j_t(\mathscr{B})=\{j_t(b),\, b \in \mathscr{B}\}$ and ${\rm vN}(\mathscr{S})$ denotes the von Neumann algebra generated by the operators in $\mathscr{S}$. Since $j_t(X)$ need not commute at different times for any $X \in \mathscr{B}$, one cannot in general assign a joint probability distribution to  the process  $\{j_t(X)\}_{t \in T}$, thus it does not have the exact analogue of a collection of finite-dimensional distributions that characterize a classical stochastic process. As a substitute for the finite-dimensional distributions, we introduce  finite-dimensional correlation kernels  $w_{\mathbf{t}_n}$. For any integer $n \geq 1$, let $\mathbf{b}_n =(b_1,b_2,\ldots,b_n) \in \mathscr{B}^n$ with $b_1,b_2,\ldots,b_n\in \mathscr{B}$ and  $\mathbf{t}_n=(t_1,t_2,\ldots,t_n) \in \mathbb{R}^n$, and define $j_{\mathbf{t}_n}(\mathbf{b}_n)=j_{t_n}(b_n) \cdots j_{t_{2}}(b_{2}) j_{t_1}(b_1)$. Then we define the finite-dimensional correlation kernels as:
\begin{equation}
w_{\mathbf{t}_n}(\mathbf{a}_n,\mathbf{b}_n) = \mu(j_{\mathbf{t}_n}(\mathbf{a}_n)^{\dag} j_{\mathbf{t}_n}(\mathbf{b}_n)). \label{eq:corr-knl}
\end{equation}
One can define a notion of equivalence between two quantum stochastic processes $(\mathscr{A}_k,\{j_{k,t}\}_{t\in T},\mu_k)$ for $k=1,2$ \cite[\S 1]{AFL82}. Also, a reconstruction theorem analogous to reconstruction theorems for stochastic processes from finite-dimensional distributions in the classical setting, such as the Kolmogorov extension theorem,  can be established \cite[Theorem 1.3]{AFL82}. Roughly speaking, it states that under certain technical assumptions on $w_{\mathbf{t}_n}$  there exists a quantum stochastic process  with  $w_{\mathbf{t}_n}$ as its correlation kernels, which is unique up to equivalence. 


Markov processes are an  important and large class of stochastic processes that are employed as models for  many applications across diverse fields. Roughly speaking, it is a process where given the present, the future of the process is independent of its past. In the quantum setting, one encounters quantum stochastic processes with an analogous property but properly interpreted since the process generally involves non-commuting operators that do not have a joint probability distribution. They are prominent in quantum optics to model a wide-range of situations involving the interaction of  localized quantum systems with travelling optical fields, and are used to accurately  describe many quantum optical devices \cite{GZ04,NY17}. 

Consider a stochastic process $\{X_t\}$ with $t \in T \subseteq \mathbb{R}$. Then the process is Markov if for any integer $n \geq 2$ and any $t_1,t_2,\ldots,t_n \in T$ satisfying $t_1<t_2<\ldots<t_n$ we have that $\mathbb{P}(X_{t_n} \in A \mid \sigma(\{X_{t_k}\}_{k=1,\ldots,n-1}))=  \mathbb{P}(X_{t_n}  \in A \mid \sigma(\{X_{t_{n-1}}\}))$ for any $A \in \mathcal{B}(\mathbb{C})$, where $\sigma(Y)$ denotes the $\sigma$-algebra generated by the random variables in $Y$. That is, given a  sequence $\{X_{t_k}\}_{k=1,2,\ldots,n-1}$, $X_{t_n}$ only depends on the most recent past, $X_{t_{n-1}}$. Due to non-commutativity, this definition cannot be extended in an analogous way to define a quantum Markov process on a quantum probability space. 

To introduce quantum Markov processes,  we follow the  treatment in \cite{AFL82}. We start with the definition of  canonical maps. 
Consider a quantum probability space $(\mathscr{A},\mu)$ and let $\mathscr{A}_{[s,t]}={\rm vN}(\{j_u(\mathscr{B}),\,s \leq  u \leq t  \in T\})$, $\mathscr{A}_{t]}=\mathscr{A}_{(-\infty,t]}$, $\mathscr{A}_t=\mathscr{A}_{[t,t]}$ and $\mathscr{A}_{[t}=\mathscr{A}_{[t,\infty)}$. A two-parameter family $\{E_{s,t}\}_{s<t \in T}$ of completely positive identity preserving maps $E_{[s,t]}: \mathscr{A}_{t]} \rightarrow \mathscr{A}_{s]}$ for $s<t \in T$, which are compatible with $\mu$ in the sense that  $
\left. \mu \right|_{\mathscr{A}_{t]}} = \left. \mu   \right|_{\mathscr{A}_{s]}} \circ E_{s,t},\,\forall s<t \in T
$, is  said to be a family of {\em canonical maps} if they have the properties:
(i) $E_{s,t}$ is a normal map and (ii) $E_{s,t}E_{t,u} = E_{s,u}$ for all $s <t< u \in T$.  We have the following definition of a quantum Markov process.
\begin{definition}[Quantum Markov process]
\item A quantum stochastic process $(\mathscr{A},\{j_t\}_{t \in T},\mu)$ over the von Neumann algebra $\mathscr{B}$ is a quantum Markov process if the canonical maps satisfy 
$E_{s,t}(\mathscr{A}_{[s,t]}) \subseteq \mathscr{A}_{s}\, \forall s<t \in T$.
\end{definition}
 The definition is slightly relaxed from the one given in \cite{AFL82} in that $\mu$ is not required to be faithful, only existence of the canonical maps is imposed.
 
Let  $j_t$ have the left inverse $j_t^*: \mathscr{A} \rightarrow \mathscr{B}$ for all $t \in T$ and define $E_{t,t}$ as the identity map on $\mathscr{A}_{t]}$.  Then we can define a two-parameter family  $\{Z_{s,t}\}_{s<t \in T}$ of completely positive identity preserving maps of $\mathscr{B}$ to itself, defined by
$
Z_{s,t} = j_s^* E_{s,t} j_t,\; s \leq t \in T
$,
satisfying
$
Z_{s,t}Z_{t,u} =Z_{s,u}\; \forall s \leq t \in T
$. In general, $\{E_{s,t}\}$ will not be conditional expectations of $\mathscr{A}_{t]}$ to $\mathscr{A}_{s]}$ unless additional technical conditions, such as from the Tomita-Takesaki theory, are satisfied \cite[p. 109]{AFL82}. Note, however, that $E_{s,t}$ will be a conditional expectation when $\mathscr{A}$ is a commutative algebra. If we assume that the canonical maps are conditional expectations then we have that \cite[Theorem 2.1]{AFL82}
\begin{eqnarray}
\lefteqn{E_{t_0,t_n}(j_{t_1}(a_1)^{\dag}\cdots j_{t_n}(a_n)^{\dag}j_t(b_n) \cdots j_{t_1}(b_1))} \notag \\
 &=&j_{t_0}Z_{t_0,t_1}(a_1^{\dag} Z_{t_1,t_2}(a_2^{\dag} \cdots Z_{t_{n-1},t_n}(a_n^{\dag}b_n)\cdots b_2)b_1),
\end{eqnarray}
for any integer $n \geq 1$, any $t_0 \leq t_1 \leq \ldots \leq t_n \in T$ and any $a_1,a_2,\ldots,a_n,b_1,b_2,\ldots,b_n \in \mathscr{B}$. When the canonical maps are also conditional expectations, there exists a one-parameter family $\{E_{t]}\}_{t \in T}$ of conditional expectations mapping $\mathscr{A}$ to $\mathscr{A}_{t]}$ for each $t \in T$, which  are compatible with $\mu$, such that $\left. E_{s,t}=E_{s]} \right|_{\mathscr{A}_{t}}$ for all $s <t \in T$. In terms of this one parameter family, the Markov property can be stated as 
$E_{s]}(\mathscr{A}_{[s})=\mathscr{A}_s,\; \forall s \in T$. The family $\{E_{s]}\}_{s \in T}$ satisfies the following properties:
\begin{eqnarray}
E_{s]}(ab) &=& E_{s]}(a)b, \forall a \in \mathscr{A},\, b \in \mathscr{A}_{s]} \label{eq:E1}\\
\mu &=& \left. \mu \right|_{\mathscr{A}_{s]} }\circ E_{s]} \label{eq:E2}\\
 E_{s]}E_{t]} &=& E_{s \wedge t]}. \label{eq:E3}
\end{eqnarray}
We can now define quantum Markov processes with conditional expectations, as follows:
\begin{definition}
\label{def:Markov-w-CE}$ (\mathscr{A},\{j_t\}_{t \in T},\mu)$ is a quantum Markov process with conditional expectations if there exists a family of normal conditional expectations $\{E_{t]}\}_{t \in T}$, mapping $\mathscr{A}$ to $\mathscr{A}_{t]}$  for each $t \in T$, satisfying $E_{s]}(\mathscr{A}_{[s})=\mathscr{A}_s\; \forall s \in T$ and \eqref{eq:E1}-\eqref{eq:E3}. 
\end{definition}

The special structure of quantum Markov processes with conditional expectations entails that their associated multi-time correlation kernels \eqref{eq:corr-knl} also have a special structure, which will be given in a theorem below. In a slightly different form (to be discussed in the next section), this result is known in the physics literature, in particular in quantum optics, as the ``quantum regression theorem".
\begin{theorem}[Quantum regression theorem]
Let $\mathbf{t}_n=(t_1,t_2,\ldots,t_n) \in T^n$ with $t_1 \leq t_2 \leq \ldots \leq t_n$. Then for any integer $n$, the time-ordered correlation kernels $w_{\mathbf{t}_n}(\mathbf{a}_n,\mathbf{b}_n)$, with $\mathbf{a}_n, \mathbf{b}_n \in \mathscr{B}^n$, of a quantum Markov process with conditional expectations are given by:
 \begin{equation}
 w_{\mathbf{t}_n}(\mathbf{a}_n,\mathbf{b}_n) = \mu \circ j_{t_1}(a_1^{\dag}Z_{t_1,t_2}(a_2^{\dag} \cdots Z_{t_{n-1},t_n}(a_n^{\dag}b_n)\cdots b_2)b_1). \label{eq:QRT1}
 \end{equation}
 \end{theorem}

 We will now make the abstract notions presented above explicit by connecting them with concrete quantum Markov models from quantum optics. We consider a localized quantum system with a finite-dimensional Hilbert space $\mathfrak{h}$ coupled to a single propagating optical field. We assume the rotating wave approximation and Markov approximation on the coupling of the system and the field \cite{GZ04}. The Hilbert space of the field is the boson (symmetric) Fock space $\mathfrak{F}=\Gamma_s(\mathcal{H})$ over the Hilbert space $\mathcal{H}=L^2(\mathbb{R}_{0+})$ of square-integrable complex functions on $\mathbb{R}_{0+}$, for details see \cite{HP84,KRP92,Mey95}.  We have the direct-sum decomposition $\mathfrak{F}=  \mathbb{C} \bigoplus \bigoplus_{k=1}^{\infty} \mathcal{H}^{\otimes_s k}$ (where $\otimes_s$ denotes the symmetric tensor product), where $\mathcal{H}^{\otimes_s k}$ is the $k$-photon subspace.  Also, for any positive integer $n$ and $t_0=0 <t_1 <t_2 < \ldots<t_n <t_{n+1}=\infty$, we have the tensor-product decomposition over time, $\mathfrak{F}  = \bigotimes_{k=0}^{n} \mathfrak{F}_{[t_k,t_{k+1}]}$, where $\mathfrak{F}_{[t_k,t_{k+1}]} =\Gamma_s(\mathcal{H}_{[t_k,t_{k+1}]})$, and $\mathcal{H}_{[t_k,t_{k+1}]}=L^2([t_k,t_{k+1}])$ is the space of square-integrable functions on $[t_k,t_{k+1}]$. 

The field has annihilation and creation densities, $b(t)$ and $b^{\dag}(t)$, respectively, that satisfy the commutation relation $[b(t),b^{\dag}(s)]=\delta(t-s)$ for all $s,t \geq 0$. For any $f \in \mathcal{H}$, let $B(f) = \int_{0}^{\infty} \overline{f(s)}b(s)ds$, $B^{\dag}(f) = \int_{0}^{\infty} f(s)b^{\dag}(s)ds$. Then we have the commutation relations, $[B(f),B(g)]=[B^{\dag}(f),B^{\dag}(g)]=0$ and $[B(f),B^{\dag}(g)]=\int_{0}^{\infty} \overline{f(s)}g(s)ds $.  Let $| \Phi \rangle$ denote  the vacuum state of the field, a state in which the field has no photons. In this state, $b(t)|\Phi \rangle=0$ for all $t \geq 0$ and thus also $B(f) |\Phi\rangle=0$. Taking $f$ informally as $f=\mathbf{1}_{[t,t+dt]}$, we can define the differentials $dB(t) = B(\mathbf{1}_{[t,t+dt]})$ and  $dB^{\dag}(t) = B^{\dag}(\mathbf{1}_{[t,t+dt]})$. It follows that  $dB(t) |\Phi \rangle=0$, $\langle \Phi |dB(t)dB^{\dag}(t)|\Phi \rangle=dt$ and $\langle \Phi |dB^{\dag}(t)dB(t)| \Phi \rangle=\langle \Phi |dB(t)dB^{\dag}(t)| \Phi \rangle=\langle \Phi |dB^{\dag}(t)dB^{\dag}(t)| \Phi \rangle=0$. This informally yields the It\={o} table in the vacuum state (see \cite{HP84,KRP92,Mey95} for a rigorous treatment):
\begin{eqnarray*}
\begin{tabular}{l|lll}
$\times $ & $dt$ & $dB$ & $dB^{\ast }$ \\ \hline
$dt$ & 0 & 0 & 0 \\ 
$dB$ & 0 & 0 &  $dt$ \\ 
$dB^{\ast }$ & 0 & 0 & 0
\end{tabular} .
\label{eq:QI_table}
\end{eqnarray*}


Take $T=\mathbb{R}_{0+}$ and let the system have Hamiltonian $H$. If the optical field is initialized in the vacuum state, the joint unitary propagator is given by the solution to a quantum stochastic differential equation (QSDE)\footnote{Here we do not consider a general QSDE as we do not include the so-called gauge or exchange process $\Lambda(t)$; see \cite{HP84} for details.} \cite{HP84}:
$$
dU(t) = (-(iH +(1/2) L^{\dag}L)dt + dB^{\dag}(t)L -L^{\dag} dB(t))U(t),
$$
with $U(0)=I$. Here $L:\mathfrak{h} \rightarrow \mathfrak{h}$ is a bounded system operator through which the system is coupled to the field. Let $\mathfrak{F}_{t]}=\mathfrak{F}_{[0,t]}$ and $\mathfrak{F}_{[t}=\mathfrak{F}_{[t,\infty)}$. Also, let $\mathtt{B}(h)$ denote the space of bounded linear operators over a Hilbert space $h$. The solution of the QSDE is  {\em adapted}, meaning that for each $t \geq 0$  $U(t) = Z(t) \otimes I_{\mathfrak{F}_{[t}}$ for some $Z(t) \in \mathtt{B}(\mathfrak{h}) \otimes \mathtt{B}(\mathfrak{F}_{t]})$, where $I_{\mathfrak{F}_{[t}}$ is the identity operator on $\mathfrak{F}_{[t}$. Note that  $\mathtt{B}(\mathfrak{h}) \otimes \mathtt{B}(\mathfrak{F}_{t]})$ is a subalgebra of $\mathtt{B}(\mathfrak{h}) \otimes \mathtt{B}(\mathfrak{F})$  by identifying any element $W \in \mathtt{B}(\mathfrak{h}) \otimes \mathtt{B}(\mathfrak{F}_{t]})$  with its ampliation $W \otimes I_{\mathfrak{F}_{[t}} \in \mathtt{B}(\mathfrak{h}) \otimes \mathtt{B}(\mathfrak{F})$.

Let $j_t(\cdot) = U(t)^{\dag} (\cdot) U(t)$.  For any system operator $X$ and $t \geq 0$, $j_t(X)$ is the evolution of $X$  in the Heisenberg picture. Now, take $\mathscr{B}= \mathtt{B}(\mathfrak{h})$ and let $\mathscr{A}$, $\mathscr{A}_{t]}$ be as defined previously (in terms of $\mathscr{B}$ and $j_t$).    By the adaptedness of $U(t)$, $\mathscr{A}_{t]}$ can be identified as a subalgebra of $\mathtt{B}(\mathfrak{h}) \otimes \mathtt{B}(\mathfrak{F}_{t]})$. We will next sketch how $(\mathscr{A},\{j_t\}_{t \in T},\mu)$ when $\mu$ is a normal state defined by $\mu(\cdot)={\rm Tr}(\rho \otimes |\Phi\rangle \langle \Phi|  \cdot )$, where  $\rho$ is the initial density operator of the system, defines a quantum Markov process with conditional expectations.  

Note that the vacuum state  has  the decomposition $|\Phi \rangle = |\Phi_{t]}\rangle |\Phi_{[t}\rangle$, with  $| \Phi_{t]} \rangle \in \mathfrak{F}_{t]}$ and $|\Phi_{[t}  \rangle \in \mathfrak{F}_{[t}$. For each $t \geq 0$, define the operator $E_t: \mathscr{A} \rightarrow \mathtt{B}(\mathfrak{h}) \otimes \mathtt{B}(\mathfrak{F}_{t]})$ via the identity $\langle \eta'  | E_t X |\eta  \rangle =  \langle \Phi_{[t} | \langle \eta' |X |\eta \rangle  |\Phi_{[t} \rangle$ for any $X \in \mathscr{A}$, and $\eta,\eta' \in \mathfrak{h} \otimes \mathfrak{F}_{t]}$ \cite[p. 214-215]{KRP92}. Via the map $E_t$ we can then define $E_{t]}:\mathscr{A} \rightarrow \mathscr{A}_{t]}$ as $E_{t]}=E_t \otimes I_{\mathfrak{F}_{[t}}$. It can be  verified that the family $\{E_{t]}\}$ satisfies \eqref{eq:E1}-\eqref{eq:E3} \cite[p. 215]{KRP92} and are thus conditional expectations. Furthermore, we can define $E_{s,t} =\left.  E_{s]} \right|_{\mathscr{A}_{t]}}$ for all $s<t \in T$ and $\{E_{s,t}\}$ is a family of canonical maps. 

In the quantum optics setting considered here, we can state the quantum regression theorem in a more explicit form, as it is usually found  in the quantum optics literature. From the definition of $E_{s]}$, it is easily verified that $E_{s]}(B(t')-B(t))=0$ for any $t'>t > s$. Using this identity and the fact that $E_{s,t}j_t=\left. E_{s]} \right|_{\mathscr{A}_{t]}} j_t = E_{s]}  j_t$, one can compute $Z_{s,t}(X)=j_s^{*}E_{s]} j_t(X)$ for any $X \in \mathtt{B}(\mathfrak{h})$ ($0\leq s \leq t$). The main step is computing the differential (with respect to $t$,  $s$ fixed), $dZ_{s,t}(X) =  j_s^{*} E_{s]}dj_t(X)$; see, e.g., \cite{HP84,KRP92,Mey95}. This yields the differential equation
$$
\frac{\partial}{\partial t} Z_{s,t}(X) = Z_{s,t}(\mathcal{L}(X)),\;0\leq s\leq t,
$$  
with initial condition $Z_{s,s}(X)=X$. Here $\mathcal{L}$ the well-known Lindblad super-operator in the Heisenberg picture given by $\mathcal{L}(X)=\frac{1}{2}L^{\dag}[X,L]+\frac{1}{2}[L^{\dag},X]L-i[X,H]$. Defining the adjoint map $Z_{s,t}^{\star}$ via the duality  ${\rm Tr}(Y Z_{s,t}(X)) ={\rm Tr}(Z_{s,t}^{\star}(Y)X)$ for all $X,Y \in \mathtt{B}(\mathfrak{h})$, it follows that $Z_{s,t}^{\star}$ satisfies the differential  equation (for fixed $s$)
$$
\frac{\partial}{\partial t} Z^{\star}_{s,t}(X) = \mathcal{L}^{\star}(Z_{s,t}^{\star}((X)),\;0\leq s\leq t,
$$  
with initial condition $Z^{\star}_{s,s}(X)=X$, where $\mathcal{L}^{\star}$ is the Lindblad super-operator in the Schr\"{o}dinger picture, as the dual to  $\mathcal{L}$,  given by $\mathcal{L}^{\star}(\rho)=\frac{1}{2}L\rho L_k^{\dag} -\frac{1}{2}(L^{\dag}L \rho + \rho L^{\dag}L)+ i[\rho,H]$. This equation has an explicit solution given by $Z_{s,t} =e^{\mathcal{L}^{\star}(t-s)}$ for $t \geq s$. In terms of $Z_{s,t}^{\star}$, the time-ordered correlation kernel \eqref{eq:QRT1} can be expressed as
\begin{equation}
w_{\mathbf{t}_n}(\mathbf{a}_n,\mathbf{b}_n) ={\rm Tr}( b_n Z_{t_{n-1},t_n}^{\star} (b_{n-1} \ldots Z_{t_2,t_3}^{\star}(b_2 Z_{t_1,t_2}^{\star}(b_1 Z_{t_1,0}^{\star}(\rho) a_1^{\dag}) a_2^{\dag})\ldots)a_{n-1}^{\dag})a_n^{\dag}). \label{eq:QRT2}
\end{equation}

The formula \eqref{eq:QRT2} is the familiar form of the quantum regression theorem found in the quantum optics literature \cite[\S 5.2]{GZ04}.  Note the  nested (or pyramidal) structure of \eqref{eq:QRT1}  and \eqref{eq:QRT2} commensurate with the time-ordering. This structure is fundamental in quantum theory and reflects the  fact that probabilities for the outcomes of sequential measurements on a quantum system (by taking the operators $a_k,b_k$  to be events in the algebra) generally depend on the order in which the measurements are performed; see \cite[\S 2.3]{GZ04} and the contribution of Gough in this volume. Also note that for a quantum Markov process with conditional expectations the time-ordered correlations is completely determined by the reduced evolution on the system (with the field traced out), as given by  \eqref{eq:QRT1} in the Heisenberg picture or \eqref{eq:QRT2} in the Schr\"{o}dinger picture.
 
\section{Summary and Future Directions}

This review article has provided a brief introduction to quantum mechanics as a quantum probability theory and  the notion of quantum stochastic processes and quantum Markov processes in the quantum probabilistic setting. From the control engineering perspective, quantum probability theory provides a theoretical foundation for the analysis and feedback control of a large class  of physical systems  with well-defined input and outputs,  which are ubiquitous in the field of quantum optics, quantum optomechanics and superconducting circuits.  Such systems are currently of interest as physical platforms for quantum technologies in sensing, communication and computing. 

Readers already familiar with probability theory can use this  existing knowledge as a basis for understanding quantum probability theory and its applications. More recent efforts in quantum stochastics include analysis of the dynamics of quantum systems interacting with travelling fields that are in highly non-Gaussian states such as single-photon states and multi-photon states, cat-states and a class of continuous matrix product states \cite{GJN11,GJN13,GJNC12,GJN14,GZ15}. An important quantum feedback operation in quantum information processing is quantum error correction \cite{Gott09,KNPM10}. Quantum feedback control will become more prominent as technology advances to a stage where more sophisticated quantum error correction codes can be experimentally demonstrated. Control of systems driven by fields in  non-Gaussian states for applications such as quantum error correction will be an interesting direction for future research. 

\section{Cross References}

Contribution of J. Gough in this volume. 
 
\section{Recommended Reading}
For a historical overview of the development of quantum probability theory see \cite{Streat00}. The text \cite{Mey95} provides an introduction to quantum probability theory for readers with a working knowledge of probability theory. Quantum stochastic calculus  was developed in \cite{HP84} and comprehensive introductions can be found in \cite{KRP92,Mey95}. The theory of quantum feedback networks based on quantum stochastic calculus can found in \cite{GJ08,GJ09}, and for an overview that emphasizes physical perspectives of the network modelling see \cite{CKS17}. For an introduction to quantum filtering theory as a quantum version of stochastic filtering theory see the contribution by Gough in this volume and \cite{BvHJ07} and the references therein. For optimal quantum feedback control and the separation principle see \cite{BvH08} and the references therein. For an introduction to linear quantum (stochastic) systems  as a quantum analog of linear stochastic systems and the modelling of  linear quantum optical devices and their applications, see \cite{NY17}.  

\bibliographystyle{IEEEtran}
\bibliography{qsp_bib}   

\end{document}